\documentclass[pdflatex,iicol]{sn-jnl}
\usepackage{amsmath,amssymb}
\usepackage{verbatim}
\usepackage{color}
\usepackage{tikz}
\usepackage{xr}
\usepackage{braket}
\usepackage[normalem]{ulem}
\usepackage{upgreek}
\usepackage[percent]{overpic}
\usepackage{graphicx,xcolor}% Include figure files
\usepackage{dcolumn}% Align table columns on decimal point
\usepackage{bm}% bold math
\usepackage[shortlabels]{enumitem}
\usepackage{bbold}
\usepackage[font=small]{caption}

\newcommand{\Tr}{\text{Tr}}
\newcommand{\Ntr}{N_{\textrm{tr}}}
\renewcommand{\rm}[1]{\textrm{#1}}
\newcommand{\Spc}{\text{Spec}}

\begin{document}

\title{Computable measures of non-Markovianity for Gaussian free fermion systems}

\author[1,2]{\fnm{Giuliano} \sur{Chiriac\`{o}}}\email{giuliano.chiriaco@dfa.unict.it}

\affil[1]{\orgdiv{Dipartimento di Fisica e Astronomia ``Ettore Majorana''}, \orgname{Universit\`{a}
di Catania}, \orgaddress{\street{via S. Sofia 64}, \city{Catania}, \postcode{95123}, \country{Italy}}}
\affil[2]{\orgname{INFN, Sez. Catania}, \postcode{I-95123}, \city{Catania}, \country{Italy}}

%\author{Giuliano Chiriac\`{o}}
%\email{giuliano.chiriaco@dfa.unict.it}
%\affiliation{Dipartimento di Fisica e Astronomia ``Ettore Majorana'', Universit\`{a} di Catania, 95123 Catania, Italy}
%\affiliation{INFN, Sez. Catania, I-95123 Catania, Italy}

%\date{\today}

%\begin{abstract}
\abstract{
We investigate measures of non-Markovianity in open quantum systems governed by Gaussian free fermionic dynamics. Standard indicators of non-Markovian behavior, such as the BLP and LFS measures, are revisited in this context. We show that for Gaussian states, trace-based distances -- specifically the Hilbert-Schmidt norm -- and second-order Rényi mutual information can be efficiently expressed in terms of two-point correlation functions, enabling practical computation even in systems where the full density matrix is intractable. Crucially, this framework remains valid even when the density matrix of the system is an average over stochastic Gaussian trajectories, yielding a non-Gaussian state. We present efficient numerical protocols based on this structure and demonstrate their feasibility through a small-scale simulation. Our approach opens a scalable path to quantifying non-Markovianity in interacting or measured fermionic systems, with applications in quantum information and non-equilibrium quantum dynamics.}
%\end{abstract}

\maketitle

\section{Introduction}

Open quantum systems inevitably interact with their surrounding environments, leading to dissipative dynamics and decoherence, thus deviating from ideal unitary evolution \cite{Breuer:Petruccione,gardinerQuantumNoise2010,fazioManyBodyOpenQuantum2024,Carmichael09}. These interactions give rise to a rich phenomenology, including non-equilibrium steady states, transient and metastable phases, and measurement-induced criticality \cite{Basov11,Lee13,nakamuraElectricfieldinducedMetalMaintained2013,chiriacoVoltageinducedMetalinsulatorTransition2018b,chiriacoPolarityDependentHeating2020b,Mitrano16,Fausti11,chiriacoTransientSuperconductivitySuperconductivity2018b,Sentef17,Kogar20,chiriacoNegativeAbsoluteConductivity2020b,Jin16,Li18,Li2019,Jian20,sierantDissipativeFloquetDynamics2022a,Sieberer13,Skinner2019}.

A common approximation assumes that such dynamics are memoryless, i.e., Markovian, and can be modeled using time-local Lindblad master equations \cite{goriniCompletelyPositiveDynamical1976,lindbladGeneratorsQuantumDynamical1976}. However, realistic physical environments often retain memory of the system’s past, resulting in a \emph{non-Markovian} behavior. The study of non-Markovian dynamics has become increasingly important due to its implications for quantum technologies, quantum communication, and quantum thermodynamics \cite{bylickaNonMarkovianityReservoirMemory2014,orieuxExperimentalOndemandRecovery2015}.

Recent literature has advanced our understanding of non-Markovianity from both physical and mathematical perspectives \cite{breuerColloquiumNonMarkovianDynamics2016b,rivasQuantumNonMarkovianityCharacterization2014,wolfAssessingNonMarkovianQuantum2008,ushadeviOpensystemQuantumDynamics2011}. Several operationally motivated measures have been introduced to quantify the degree of non-Markovianity in a dynamical process. Among the most widely used are the Breuer–Laine–Piilo (BLP) measure \cite{breuerMeasureDegreeNonMarkovian2009}, which is based on the backflow of information as seen in trace distance revivals; the Rivas–Huelga–Plenio (RHP) measure \cite{rivasEntanglementNonMarkovianityQuantum2010b}, which is based on the divisibility of the quantum map; and the Luo–Fu–Song (LFS) measure \cite{luoQuantifyingNonMarkovianityCorrelations2012}, which detects memory effects via increases in the mutual information between the system and an ancilla.

While these measures are well defined in general, their computation becomes prohibitive in large many-body systems, since it requires evolving full density matrices and optimizing over initial conditions -- a task that scales exponentially with system size. To address this issue, we focus on a class of open quantum systems, namely Gaussian free fermionic systems \cite{fagottiEntanglementEntropyTwo2010a,suraceFermionicGaussianStates2022c}. The state of a Gaussian system is fully characterized by its two-point correlation functions,  allowing us to reformulate key non-Markovianity measures in terms of correlation matrices alone.

In this work, we propose a new method to calculate numerically non-Markovianity measures, test it against existing methods and show that it reduces the computational cost from exponential to polynomial in system size.

In particular, we show that: \emph{(i)} the Hilbert-Schmidt distance can replace the trace norm in the BLP measure and be computed directly from correlation functions. \emph{(ii)} A variant of the LFS measure based on second-order Rényi entropy can likewise be computed from correlation matrices. Our methods leverage quantum trajectory simulations of Lindbladian evolutions that preserve Gaussianity, reducing the computational cost from exponential to polynomial scaling in system size and trajectory count.

This technique is particularly appealing for dissipative free fermionic systems, which have been extensively studied in recent year in the context of measurement induced transitions \cite{caoEntanglementFermionChain2019a,coppolaGrowthEntanglementEntropy2022a,tsitsishviliMeasurementInducedTransitions2024b,poboikoTheoryFreeFermions2023,starchlGeneralizedZenoEffect2025,leungTheoryFreeFermions2023,szyniszewskiDisorderedMonitoredFree2023,kellsTopologicalTransitionsWeakly2023,albertonEntanglementTransitionMonitored2021a,poboikoMeasurementinducedTransitionsInteracting2025,muzziEntanglementEnhancementInduced2025,piccittoEntanglementDynamicsString2023,russomannoEntanglementTransitionsQuantum2023,favaNonlinearSigmaModels2023}

As a simple benchmark, we simulate a free fermionic chain of $L$ sites with dephasing processes, engineered to produce non-Markovian dynamics via coupling to a finite auxiliary system. We compute the BLP measure using both Gaussian-preserving quantum trajectories and two-point correlation matrices methods, and full-state evolution methods (with QuTiP), and compare the computational cost of both methods for different values of $L$, showing that the Gaussian method exhibits an exponential advantage at large system sizes.

The rest of the paper is structured as follows. In Section \ref{Sec:Preliminaries} we introduce the model and review the properties of free fermionic systems. In Section \ref{sec:NMmeasures} we review the most famous non-Markovianity measures in the literature. In Section \ref{sec:NM-Gaussian} we show how to compute such measures for Gaussian preserving dynamics and outline a protocol for the calculation of the BLP and LFS measures. In Section \ref{Sec:numerics} we present the results of our numerical simulations and compare the efficiency of our protocol compared to a full state numerical simulations. Finally in Section \ref{sec:Conclusions} we present our conclusions and outlook.

\begin{figure*}[t!]
    \setlength{\unitlength}{1cm}
    \begin{picture}(0,0)
    \put(0.2,2.8){a)}
    \put(6.85,2.8){b)}
    \put(11.95,2.8){c)}
    \end{picture} 
    \includegraphics[width=\textwidth]{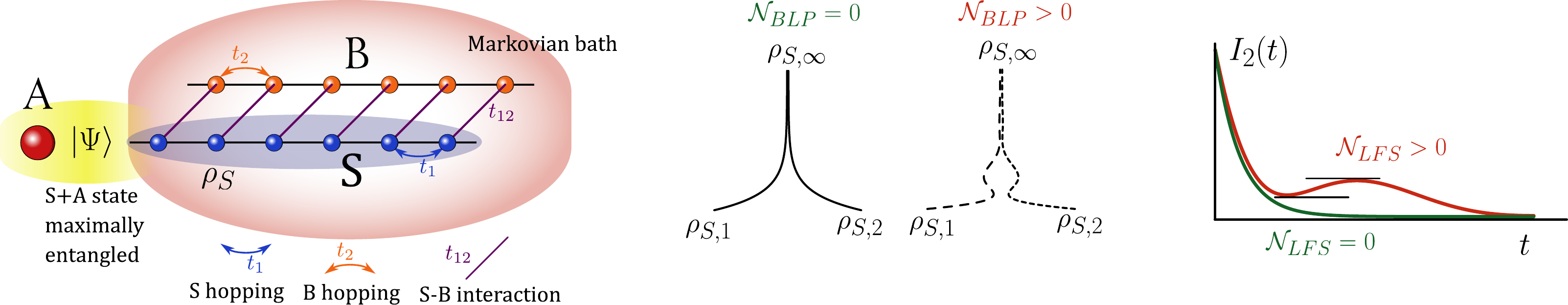}
    \caption{(a) Sketch of the system under consideration. The system $S$ is composed by fermionic sites along which the electrons can hop with strength $t_1$. The ancillary degrees of freedom $B$ are also given by fermions that hop with strength $t_2$. $S$ and $B$ are coupled by fermionic hopping $t_{12}$ and they undergo a total Markovian dynamics. When the $LFS$ non-Markovianity measure is considered, an ancilla $A$ is coupled to the system, and $S+A$ is initialized in a completely entangled state $\ket{\Psi}$. (b) The BLP non-Markovianity measure studies the time evolution of two different initial conditions $\rho_{S,1}$ and $\rho_{S,2}$, which converge to the same steady state $\rho_{S,\infty}$. When the dynamics is Markovian, the distance between the density matrices decay in a monotonic way and $\mathcal{N}_{BLP}=0$ (left). On the other hand when the dynamics is non-Markovian the two density matrices do not converge monotonically to the steady state and $\mathcal{N}_{BLP}>0$ (right). (c) The LFS non-Markovianity measure studies the time evolution of the mutual information $I_2$ between the system and the ancilla $A$. When $I_2$ decays monotonicly (green curve) the system is Markovian and $\mathcal{N}_{LFS}=0$. When $I_2$ exhibits times where it increases, then the system is non-Markovian and $\mathcal{N}_{LFS}>0$ (red curve).}
    \label{Fig:sketch}
\end{figure*}
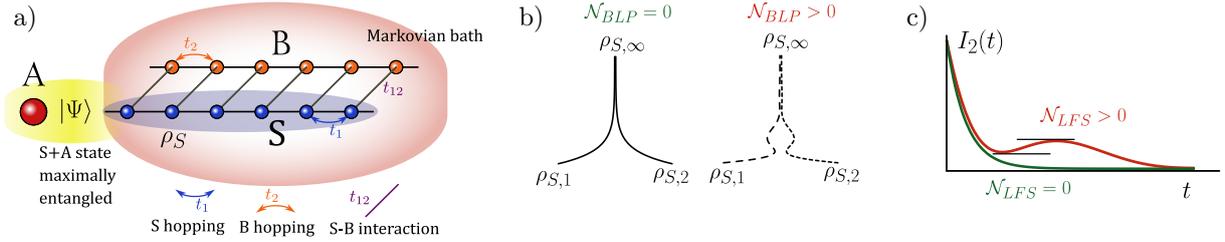

\section{The model}
\label{Sec:Preliminaries}

We are concerned with studying the evolution of a non-Markovian free fermionic system, whose density matrix evolves according to a map $\Phi_t$, with $t$ the time, such that the density matrix $\rho$ is
\begin{equation}\label{eq:rho_NM}
\rho(t)=\Phi_t(\rho(0))
\end{equation}

Several classes of non-Markovian dynamics can be treated by adding some suitable degrees of freedom (which we indicate with $B$) to the system in such a way that the dynamics of the extended system is Markovian. As an example, this can always be done for systems whose dynamics is described by a time-local master equation in a TCL form, as outlined in \cite{breuerGenuineQuantumTrajectories2004a}. We assume that our non-Markovian map $\Phi$ can be treated in this way, so that the density matrix of the system $S$ plus the ancillary degrees of freedom $S+B$ evolves according to a Lindblad master equation:
\begin{equation}
\dot\rho^{S+B}=\mathcal{L}\rho^{S+B}; \quad \rho^{S+B}(t)=\Phi_t^{S+B}(\rho^{S+B}(0))
\label{eq:Lindblad}
\end{equation}
where $\Phi_t^{S+B}=e^{\mathcal{L}t}$. We are interested in the reduced density matrix of the system $\rho(t)=\rho^S(t)=\text{Tr}_B\rho^{S+B}(t)$.

Since the dynamics of system+ancilla is Markovian, it can be simulated using the quantum trajectories technique \cite{Daley2014,Plenio98,molmerMonteCarloWavefunction1993,dalibardWavefunctionApproachDissipative1992}, where each trajectory corresponds to a random realization of the Kraus operators that describe the quantum jumps. Formally
\begin{equation}\label{eq:rho_QTraj}
\rho^{S+B}(t)\approx\frac1{\Ntr}\sum_{\alpha=1}^{\Ntr}\rho_{\alpha}^{S+B}(t)
\end{equation}
where $\alpha$ labels the trajectory, $\Ntr$ is the total number of trajectories used in the numerical simulations, $\rho_{\alpha}^{S+B}=\ket{\psi_{\alpha,S+B}}\bra{\psi_{\alpha}^{S+B}}$ is the density matrix in trajectory $\alpha$, corresponding to the pure state $\ket{\psi_{\alpha}^{S+B}}$.

We consider dynamics that preserve the Gaussianity of the quantum state. A Gaussian state can be fully described in terms of two-point correlation functions, as we will see in detail in the next subsection. Examples of such dynamics, and how it is simulated using quantum trajectories are given in \cite{tsitsishviliMeasurementInducedTransitions2024b,muzziEntanglementEnhancementInduced2025}. In such cases, each trajectory can be simulated efficiently, in a time that scales polynomially with the size of the system, so that the complexity is $\sim\mathcal{O}(L^a\Ntr)$, with $a\sim2\div3$ a coefficient dependent on the efficiency the implementation of matrix multiplication, opposed to the complexity $\sim\mathcal{O}(2^{2L})$ of an exact simulation.

\subsection{Gaussian fermionic states}
\label{Sec:GaussianStates}

We consider a free fermionic open quantum system with $L$ sites, described by canonical fermionic creation and annihilation operators $c_n^\dagger$ and $c_n$. It is often convenient to work with Majorana fermions, defined as:
\begin{equation}
a_{2n-1} = c_n + c_n^\dagger, \qquad a_{2n} = -i(c_n - c_n^\dagger)
\label{eq:Majorana}
\end{equation}
where $a_j$ are Majorana operators which satisfy $\{a_j, a_k\} = 2\delta_{jk}, \quad a_j^\dagger = a_j$.

A fermionic state is Gaussian if it can be completely characterized by its two-point correlation functions. We consider two commonly used correlation matrices:
\begin{gather}
C_{nm} = \text{Tr}(\rho c_n^\dagger c_m) = \langle c_n^\dagger c_m \rangle;
\label{eq:Ccorr} \\
\Gamma_{jk} = \frac{i}{2} \text{Tr}(\rho [a_j, a_k]) = i \left( \langle a_j a_k \rangle - \delta_{jk} \right)
\label{eq:GammaCorr}
\end{gather}

The first equation defines the fermionic correlation matrix, a $L\times L$ Hermitian matrix; while the second line defines the Majorana correlation matrix, a $2L \times 2L$ real anti-Hermitian matrix. If the number of particles is conserved during the dynamics of the system\footnote{If the dynamics does not conserve the number of particles, a second correlation matrix $F_{nm} = \langle c_n c_m \rangle$ needs to be introduced.}, the two matrices are related, and the Majorana correlations can be written in terms of the fermionic correlations as:
\begin{equation}
\Gamma = (2C - \mathbb{1})\otimes \begin{pmatrix} 0 & i \\ -i& 0 \end{pmatrix}
\label{eq:GammaFromC}
\end{equation}

Within this formalism, the calculation of several quantities -- e.g. partial traces, entropies, eigenvalues -- can be performed efficiently. The density matrix of a Gaussian state can be written as the exponential of
\begin{equation}
\rho = \frac{\exp\left(\frac{1}{4} a W a \right)}{Z(W)}
\label{eq:rhoGaussian}
\end{equation}
where \(a W a = \sum_{ij} a_i W_{ij} a_j\), and \(Z(W) = \mathrm{Tr}\left[ \exp\left( \frac{1}{4} a W a \right) \right]\) is the normalization factor, often referred to as the "partition function" of the Gaussian state. The matrix \(W\) is antisymmetric and is related to the Majorana correlation matrix of the system by
\begin{equation}
\Gamma=\tanh\left(\frac{W}{2}\right);\qquad \exp(W) = \frac{1 + \Gamma}{1 - \Gamma}
\label{eq:GammaTanh}
\end{equation}

Since $W$ is anti-symmetric, its eigenvalues come in pairs $+\lambda_w,-\lambda_w$. The Gaussian partition function can be expressed as \cite{fagottiEntanglementEntropyTwo2010a}
\begin{equation}
Z(W) = \sqrt{\prod_{\lambda_w} 2\cosh\left( \frac{\lambda_w}{2} \right)}
\label{eq:Zcosh}
\end{equation}

or alternatively%, in terms of the eigenvalues \(\lambda_{e^W}\) of \(e^W\), as:
\begin{equation}
Z(W) = \sqrt{\prod_{\lambda_{w}} \left( \sqrt{e^{\lambda_{w}}} + \frac{1}{\sqrt{e^{\lambda_{w}}}} \right)}.
\label{eq:ZexpW}
\end{equation}

When the state of the system is a statistical mixture over several Gaussian trajectories, such as in Eq. \eqref{eq:rho_QTraj}, the density matrix is expressed as:
\begin{equation}
\rho = \frac{1}{N_{\mathrm{tr}}} \sum_{\alpha=1}^{N_{\mathrm{tr}}} \frac{\exp\left(\frac{1}{4} a W_{\alpha} a \right)}{Z(W_{\alpha})},
\label{eq:rhoMixedGaussian}
\end{equation}
where $W_\alpha$ dscribes the density matrix of the $\alpha$-th Gaussian trajectory: $\rho_\alpha=\exp(\frac{1}{4} a W_{\alpha} a)/Z(W_\alpha)$. We note that the density matrix in Eq. \eqref{eq:rhoMixedGaussian} is no longer a Gaussian state. However, several quantities of interest can still be calculated from the knowledge of $W_\alpha$, as we will see in the next section for the case of the quadratic trace distance.

\section{Non-Markovianity measures}
\label{sec:NMmeasures}

The characterization and quantification of non-Markovianity in open quantum systems require a more nuanced analysis due to the dependence on the history of the system. To that end, a number of measures have been introduced to quantify the degree of non-Markovianity in a dynamical map; each reflects a different operational or information-theoretic aspect of the dynamics~\cite{breuerColloquiumNonMarkovianDynamics2016b,rivasQuantumNonMarkovianityCharacterization2014}. Here we summarize the definitions and physical principles behind the most common non-Markovianity measures.

\paragraph{Information backflow: BLP measure} A widely used approach, introduced by Breuer, Laine, and Piilo (BLP)~\cite{breuerMeasureDegreeNonMarkovian2009}, is based on the concept of information backflow. The BLP measure quantifies non-Markovianity by monitoring the trace distance between two quantum states $\rho_p(t)$ and $\rho_q(t)$ evolving under the same dynamical map:
\begin{equation}
\label{eq:BLP-tracedist}
d_1(\rho_p(t), \rho_q(t)) = \frac{1}{2} \|\rho_p(t) - \rho_q(t)\|_1,
\end{equation}
where $\|A\|_1 \equiv \text{Tr}(\sqrt{A^{\dagger}A})$ denotes the trace norm. In a Markovian evolution, the trace distance is contractive, meaning that it can only decrease over time:
\begin{equation}
\label{eq:BLP-contractivity}
d_1(\Phi_t(\rho_p), \Phi_t(\rho_q)) \leq d_1(\rho_p, \rho_q)
\end{equation}
for any completely positive and trace-preserving (CPTP) map $\Phi_t$. Thus, any temporary increase in trace distance can be interpreted as a signature of information flowing back from the environment to the system, and hence non-Markovianity.

The BLP measure is defined as
\begin{equation}
\label{eq:BLP-measure}
\mathcal{N}_{\text{BLP}} = \max_{\rho_{p,q}(0)} \int_{\dot d_1 > 0} dt \; \partial_t d_1(\rho_p(t), \rho_q(t)).
\end{equation}
The maximization is typically restricted to orthogonal pure states. While this measure has an appealing operational interpretation, its computation can be challenging due to the required maximization over all pairs of initial states.

We note that other distance measures can be used in place of $d_1$, provided they satisfy contractivity under CPTP maps (see Appendix~\ref{Appendix:d2}). Two notable examples are the Hilbert-Schmidt distance $d_2$ and the quantum relative entropy $d_{\mathrm{RE}}$:
\begin{align}
\label{eq:d2-def}
d_2(\rho_p, \rho_q) &\equiv \sqrt{\frac{1}{2} \mathrm{Tr}|\rho_p - \rho_q|^2} \\
\label{eq:dRE-def}
d_{\mathrm{RE}}(\rho_p, \rho_q) &\equiv \mathrm{Tr}[\rho_p(\ln\rho_p - \ln\rho_q)]
\end{align}
In particular, we will see in Sec.~\ref{sec:NM-Gaussian} that $d_2$ can be efficiently computed for Gaussian fermionic states.

\paragraph{Correlations revival: LFS measure} Another measure, introduced by Luo, Fu, and Song (LFS)~\cite{luoQuantifyingNonMarkovianityCorrelations2012}, builds upon the dynamics of correlations between the system and an ancilla $A$ %\footnote{we call it $A'$ to distinguish it from the ancillay degrees of freedom used to make the total dynamics Markovian}
. Given an initial system-ancilla state $\rho_{S+A}(0)$, where the system evolves under the map $\Phi_t$ and the ancilla remains unchanged, correlations are quantified via the mutual information:
\begin{equation}
\label{eq:mutual-info}
I(S:A) = S_{vN}(\rho_S) + S_{vN}(\rho_{A}) - S_{vN}(\rho_{S+A}),
\end{equation}
where $S_{vN}(\rho)\equiv-\Tr(\rho\ln\rho)$ denotes the von Neumann entropy. The mutual information $I$ decreases under local CPTP maps. Conversely, an increase in $I$ signals a non-Markovian process due to a revival of correlations. The LFS measure quantifies non-Markovianity as the total amount of revival in mutual information, maximized over initial system-ancilla states:
\begin{equation}
\label{eq:LFS}
\mathcal{N}_{LFS}\equiv\max_{\rho^{S+A}(0)} \int_{\dot I > 0} dt \; \partial_t I(t).
\end{equation}

\paragraph{CP-Divisibility: RHP measure} A conceptually distinct notion of Markovianity is based on the divisibility of the dynamical map, as outlined by Rivas, Huelga, and Plenio (RHP) in \cite{rivasEntanglementNonMarkovianityQuantum2010b}. We do not investigate the behavior of this measure.

\section{Non-Markovianity in Gaussian fermionic systems}
\label{sec:NM-Gaussian}

The study of non-Markovianity in Gaussian systems has been the subject of constant work, via witnesses based fidelity, quantum state distance, and so on \cite{scutaruFidelityDisplacedSqueezed1998,marianDistinguishabilityNonclassicalityOnemode2004,vasileQuantifyingNonMarkovianityContinuousvariable2011,frigerioExploitingGaussianSteering2021,mouloudakisNonMarkovianityTimeEvolution2023,hesabiNonMarkovianityMeasureGaussian2021}. 

The calculation of non-Markovianity measures can be efficiently performed for Gaussian fermionic states, when the relevant quantities can be reduced to functions of two-point correlation matrices. However, care must be taken when the system state is a convex mixture of Gaussian states -- e.g. the density matrix arising from the average over quantum trajectories -- which results in a non-Gaussian state for which the trace distances or entropies employed in the calculation of non-Markovianity measures may not be directly expressed in terms of the correlation matrices.

\subsection{BLP Measure}
\label{sec:BLP}

Let us consider the case of the trace norm $d_1(\rho_p(t),\rho_q(t))=\frac12\Tr\sqrt{|\rho_p(t)-\rho_q(t)|^2}$ vs the Hilbert-Schmidt norm $d_2(\rho_p(t),\rho_q(t))=\sqrt{\frac12\Tr|\rho_p(t)-\rho_q(t)|^2}$, employed in the calculation of the BLP measure.

We write the density matrix as an average over quantum trajectories $\rho_p(t) = \frac{1}{\Ntr} \sum_{\alpha=1}^{\Ntr}\rho_p^{\alpha(t)}$:
\begin{equation}
\label{eq:rho_traj_avg}
\rho_{p}(t) =\frac{1}{\Ntr} \sum_{\alpha=1}^{\Ntr} \frac{\exp\left(\frac{1}{4} a W_p^{\alpha}(t) a \right)}{Z(W_p^{\alpha}(t))}
\end{equation}
and similarly for \(\rho_q(t)\). The key observation is that the product of two Gaussian states is again a Gaussian state \cite{fagottiEntanglementEntropyTwo2010a}:
\begin{equation}
\label{eq:GaussianProduct}
e^{\frac{1}{4} a W a} e^{\frac{1}{4} a W' a} = e^{\frac{1}{4} a (W \oplus W') a}, \quad e^{W \oplus W'} \equiv e^W e^{W'}.
\end{equation}

Therefore each term in the sum $|\rho_p- \rho_q|^2=(\rho_p- \rho_q)^2$ is Gaussian, and we can write
\begin{gather}
\label{eq:dist_pq}
|\rho_p- \rho_q|^2=\frac1{\Ntr^2}\sum_{\alpha,\beta}\Big(\frac{e^{\frac{1}{4} a W_{p,p}^{\alpha+\beta} a}}{Z(W_p^{\alpha}) Z(W_p^{\beta})}+\\
\notag+\frac{e^{\frac{1}{4} a W_{q,q}^{\alpha+\beta} a}}{Z(W_q^{\alpha}) Z(W_q^{\beta})}+\frac{e^{\frac{1}{4} a W_{p,q}^{\alpha+\beta} a}+e^{\frac{1}{4} a W_{q,p}^{\beta+\alpha} a}}{Z(W_p^{\alpha}) Z(W_q^{\beta})}\Big);\\
\label{eq:Trdist_pq}
\Tr|\rho_p- \rho_q|^2=\frac1{\Ntr^2}\sum_{\alpha,\beta}\Big(\frac{Z(W_{p,p}^{\alpha+\beta})}{Z(W_p^{\alpha}) Z(W_p^{\beta})}+\\
\notag+\frac{Z(W_{q,q}^{\alpha+\beta})}{Z(W_q^{\alpha}) Z(W_q^{\beta})}+\frac{2Z(W_{p,q}^{\alpha+\beta})}{Z(W_p^{\alpha}) Z(W_q^{\beta})}\Big),
\end{gather}
where $W_{p,q}^{\alpha+\beta}\equiv W_p^\alpha\oplus W_q^\beta$.

From Eq. \eqref{eq:dist_pq} we note that the sum is not Gaussian in general, and thus evaluating $\sqrt{|\rho_p-\rho_q|^2}$ -- which is needed for the calculation of $d_1$ -- requires expanding in series and evaluating an infinite number of terms. This makes the numerical calculation of $d_1$ extremely inefficient, if not impossible. The same difficulty is encountered when calculating $d_{RE}$, since the calculation of $\ln\rho_{p,q}$ requires an expansion in series with infinite terms.

On the other hand, Eq. \eqref{eq:Trdist_pq} shows that the trace of $|\rho_p-\rho_q|^2$ can be easily expressed in terms of the partition functions of each Gaussian trajectory.

Therefore, a variation of the BLP measure based on $d_2$
\begin{equation}
\label{eq:N2BLP}
\mathcal{N}_{\text{BLP},2} = \max_{\rho_{p,q}(0)} \int_{\dot d_2 > 0} \!\! dt \; \partial_t d_2(\rho_p(t), \rho_q(t)).
\end{equation}
can be efficiently computed from the correlation matrix in a computational time that scales polynomially with the system size.

We can thus outline a protocol to calculate $\mathcal{N}_{\text{BLP},2}$ for a system $S$ undergoing an evolution governed by the map $\Phi_t$ that preserves Gaussianity. We assume that we can extend the Hilbert space, adding an ancilla $A$ such that the total dynamics is described by the map $\Phi_t^{S+B}$ which is Markovian and can be simulated via (Gaussianity-preserving) quantum trajectories. Then, the protocol is

\emph{(i)}. We first choose two initial Gaussian states \( \rho_{p,q}^{S+B} \), corresponding to the correlation matrices \( C^{\alpha,S+B}_{p,q}(0) \). Since we only need to maximize over the system initial states, we can choose them to be factorized states \( \rho_{p,q}^{S+B} = \rho_{p,q}(0)\otimes\rho^B(0) \).

\emph{(ii)}. We evolve the system along a quantum trajectory $\alpha$ and obtain \( C^{\alpha,S+B}_{p,q}(t) \), describing the Gaussian density matrix \( \rho^{\alpha,S+B}_{p,q}(t) \). The average over trajectories $\alpha$ of this density matrix gives \(\rho_{p,q}^{S+B}(t)=\Phi_t^{S+B}(\rho^{S+B}_{p,q}(0))\)

\emph{(iii)}. We trace out the ancilla degrees of freedom, by constructing the reduced correlation matrix \( C^{\alpha}_{p,q}(t) \), which describes the density matrix of the system for trajectory $\alpha$: \( \rho^{\alpha}_{p,q}(t)\). The average over trajectories of this density matrix gives the state of the system at time $t$: \( \rho_{p,q}(t)=\Phi_t(\rho_{p,q}(0))\)

\emph{(iv)}. From \( C^{\alpha}_{p,q}(t) \) we first calculate \(\Gamma_{p,q}^\alpha(t)\) from Eq. \eqref{eq:GammaFromC}, then \(W_{p,q}^\alpha(t)\) from Eq. \eqref{eq:GammaTanh}, and finally \(Z(W_{p,q}^\alpha(t)) \) from Eq. \eqref{eq:ZexpW}.

\emph{(v)}. We compute $d_2$ using Eqs. \eqref{eq:d2-def} and \eqref{eq:Trdist_pq}.

\emph{(vi)}. Using Eq. \eqref{eq:N2BLP} we calculate \( \mathcal{N}_{\text{BLP},2} \).

Step \emph{(ii)} of the protocol can be done in a time that usually scales as $\mathcal{O}(\Ntr\cdot N_t\cdot L^3)$ where $N_t$ is the number of time steps required for each quantum trajectory. On the other hand, steps \emph{(iv)}-\emph{(v)} can be done in a computational time $\sim\mathcal{O}(\Ntr^2\cdot L^3)$, since the calculation of the eigenvalues of an $L\times L$ matrix generally scales as $L^3$. Since typically $\Ntr\gtrsim N_t$, the complexity of the algorithm is then $\sim\mathcal{O}(\Ntr^2L^3)$. Therefore, the complexity of calculating the BLP non-Markovianity measure scales polynomially with both the size of the system and the number of trajectories. On the other hand, simulating the dynamics of the full quantum state of the system, even using the quantum trajectories approach, scales at least with a complexity $\sim\mathcal{O}(\Ntr N_t2^L)$, which is exponentially larger than the complexity of the Gaussian non-Markovianity measure.

\subsection{LFS Measure}
\label{sec:LFS}

A similar argument can be made for the calculation of the LFS measure. If the dynamics of the system preserves Gaussianity, then adding an ancilla $A$ with identity dynamics also preserves Gaussianity, so that the dynamics of $S+A$ can still be simulated via quantum trajectories evolving only the correlation matrix $C^{\alpha,S+A+B}(t)$. After tracing out the degrees of freedom of $B$, the density matrix of $S+A$ can be written as
\begin{equation}
\label{eq:rho_traj_avg_LFS}
\rho^{S+A}(t) = \frac{1}{N_{\mathrm{tr}}} \sum_{\alpha=1}^{N_{\mathrm{tr}}} \frac{\exp\left(\frac{1}{4} a W^{\alpha,S+A}(t) a \right)}{Z(W^{\alpha,S+A}(t))},
\end{equation}
where $W^{\alpha,S+A}(t)$ is obtained from the correlation matrix reduced $C^{\alpha,S+A}(t)$ to the $S+A$ degrees of freedom.

In order to obtain the reduced density matrices $\rho^S(t)$ and $\rho^{A'}(t)$ we just need to further reduce $C^{\alpha,S+A}(t)$, by tracing out either $A$ or $S$. However, if we want to calculate the von-Neumann entropy of $\rho^{S+A}(t)$ using the correlation matrix of each trajectory, we need to expand in series the logarithm, which means evaluating an infinite number of terms, a task which is numerically impossible. This is the same difficulty encountered for the numerical evaluation of $d_1$ and $d_{RE}$. 

However, we can still define a mutual information based on the second-Renyi entropy $S_2(\rho)\equiv\Tr\rho^2$:
\begin{gather}
\label{eq:I2_def}
I_2(S:A)\equiv S_{2}(\rho_S) + S_{2}(\rho_{A}) - S_{2}(\rho_{S+A});\\
\label{eq:N2LFS}
\mathcal{N}_{LFS,2}\equiv\max_{\rho^{S+A}(0)} \int_{\dot I_2 > 0} dt \; \partial_t I_2(t).
\end{gather}

Since we can write $I_2=\Tr(\rho_S^2+\rho_{A}^2-\rho_{S+A}^2)$
%
%\begin{figure*}[h!]
\begin{gather}
\label{eq:I2_sum}
I_2=\frac1{\Ntr^2}\sum_{\alpha,\beta}\Big(\frac{Z(W^{S,\alpha+\beta})}{Z(W^{S,\alpha}) Z(W^{S,\beta})}+\\
\notag+\frac{Z(W^{A,\alpha+\beta})}{Z(W^{A,\alpha}) Z(W^{A,\beta})}-\frac{Z(W^{S+A,\alpha+\beta})}{Z(W^{S+A,\alpha}) Z(W^{S+A,\beta})}\Big),
\end{gather}
%\end{figure*}%\twocolumn
%
the calculation of $I_2$, and thus of $\mathcal{N}_{LFS,2}$ can be performed in a polynomial time that scales again as $\sim\mathcal{O}(\Ntr^2L^3)$.

The protocol to calculate the LFS measure for Gaussian states can then be outlined as

\emph{(i)}. We choose the initial Gaussian states \( \rho^{S+A+B} \), corresponding to the correlation matrices \( C^{\alpha,S+A+B}(0) \). We can choose the $S+B$ state to be factorized with $A$: \( \rho^{S+A+B}(0) = \rho^{S+A}(0)\otimes\rho^B(0) \).

\emph{(ii)}. We evolve the system along a quantum trajectory $\alpha$ and obtain \( C^{\alpha,S+A+B}(t) \), describing the Gaussian density matrix \( \rho^{\alpha,S+A+B}(t) \). The average over trajectories $\alpha$ of this density matrix gives \(\rho^{S+A+B}(t)=\Phi_t^{S+B}\otimes\mathbb{1}_{A}(\rho^{S+A+B}(0))\)

\emph{(iii)}. We trace out the $A$ degrees of freedom, by constructing the reduced correlation matrix \( C^{\alpha,S+A}(t) \), which describes the density matrix of $S+A$ for trajectory $\alpha$: \( \rho^{\alpha,S+A}(t)\). The average over trajectories of this density matrix gives the state of the system at time $t$: \( \rho^{S+A}(t)=\Phi_t\otimes\mathbb{1}_{A}(\rho^{S+A}(0)) \)

\emph{(iv)}. From \( C^{\alpha,S+A}(t) \) we calculate the reduced density matrices \( C^{\alpha,S}(t) \) and \( C^{\alpha,A}(t) \). We then first evaluate \(\Gamma^{\alpha,S+A}(t)\), \(\Gamma^{\alpha,S}(t)\), \(\Gamma^{\alpha,A}(t)\) from Eq. \eqref{eq:GammaFromC}, then \(W^{\alpha,S+A}(t)\), \(W^{\alpha,S}(t)\), \(W^{\alpha,A}(t)\) from Eq. \eqref{eq:GammaTanh}, and finally \(Z(W^{\alpha,S+A}(t)) \), \(Z(W^{\alpha,S}(t)) \), \(Z(W^{\alpha,A}(t)) \) from Eq. \eqref{eq:ZexpW}.

\emph{(v)}. We compute $I_2$ using Eq. \eqref{eq:I2_sum}.

\emph{(vi)}. Using Eq. \eqref{eq:N2BLP} we calculate \( \mathcal{N}_{\text{LFS},2}\).

\section{Numerical tests}\label{Sec:numerics}

To illustrate and benchmark our method, we numerically evaluate the non-Markovianity measure \( \mathcal{N}_{\mathrm{BLP},2} \) using both the Gaussian trajectory approach developed in this work and a full density matrix simulation of the dynamics of the system. The latter is implemented using the QuTiP library \cite{johanssonQuTiPOpensourcePython2012,johanssonQuTiP2Python2013,lambertQuTiP5Quantum2024}, which allows direct integration of Lindblad master equations but scales exponentially with system size.

\begin{figure*}[t!]
    \centering
    \includegraphics[width=\linewidth]{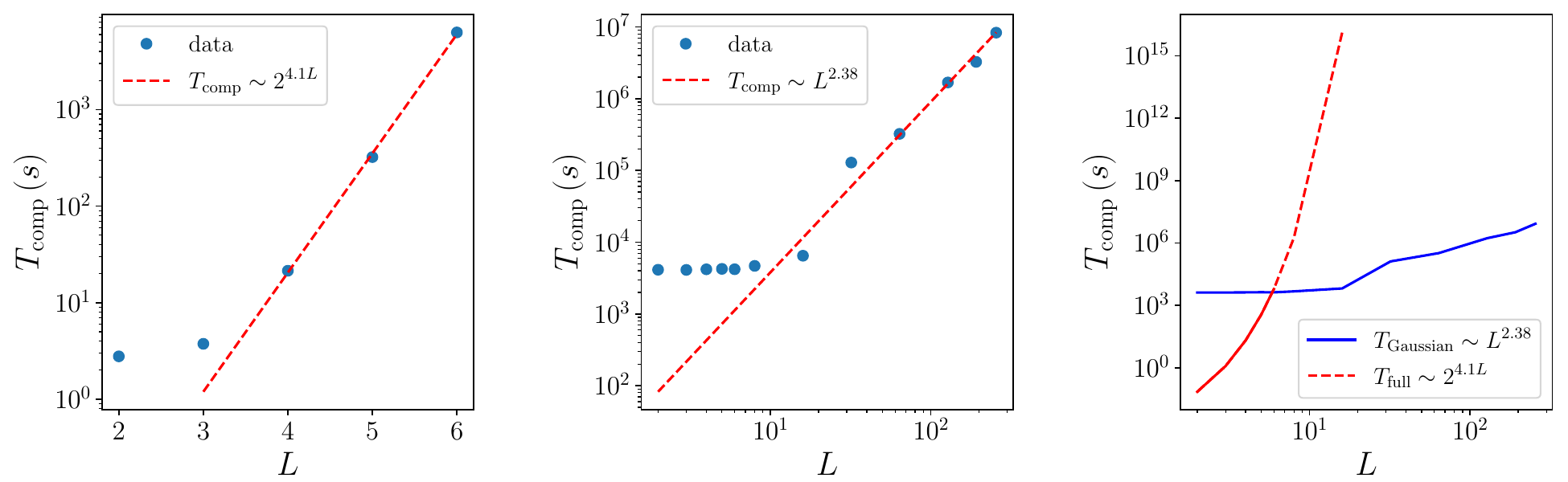}
    \setlength{\unitlength}{1cm}
    \begin{picture}(0,0)
    \put(-7.8,4.9){a)}
    \put(-2.45,4.9){b)}
    \put(3,4.9){c)}
    \end{picture}
    \caption{Plots of the computational time required to calculate $d_2$ for a system with $L$ fermionic sites (plus $L$ sites for the ancillary degrees of freedom). a) Computational time required for a simulation of the full quantum state using QuTiP for $L=2,3,4,5,6$; the blue points indicate the data obtained from the simulations, while the red dashed line indicates an exponential fit to the data, from which we estimate $T_{\mathrm{comp}}\sim2^{4.1L}$, in line with the theoretical prediction $\sim2^{2(2L)}\sim2^{4L}$. For $L=2,3$ the computational time is dominated by the overhead of the algorithm, while for larger $L$ the exponential increase is very evident. b) Computational time required for a simulation using quantum trajectories for Gaussian states for $L$  ranging between $L=2$ and $L=256$; the blue points indicate the data obtained from the simulations, while the red dashed line indicates a power-law fit to the data, from which we estimate $T_{\mathrm{comp}}\sim L^{2.38}$. Also in this case, $T_{\mathrm{comp}}$ is dominated by the computational overhead at small system sizes, and then follows a power-law behavior. c) Comparison of $T_{\mathrm{comp}}$ for the two methods. The Gaussian method has a larger overhead then the full state simulation method, which makes computationally more costly at low $L$. However, already for $L\approx6$, the Gaussian method requires less resources; for $L\gtrsim10$ the Gaussian method is the only one feasible.}
    \label{fig:blp_comparison}
\end{figure*}
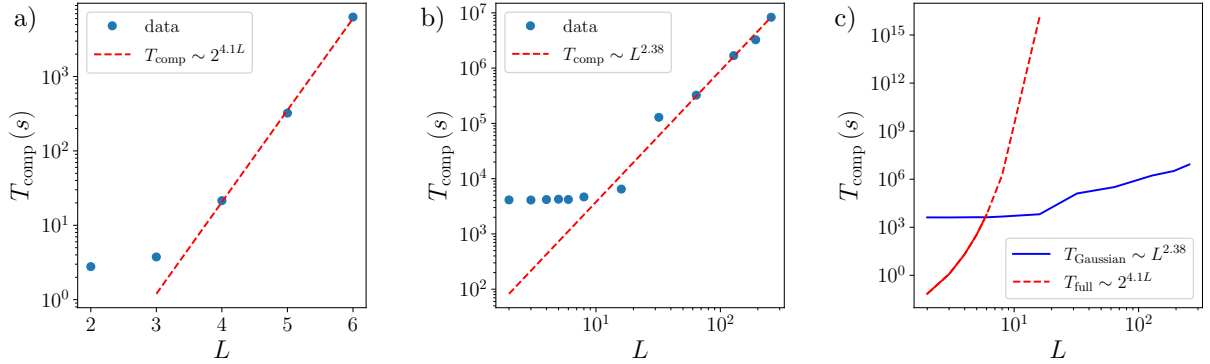

We consider a two-chain free fermionic system, each with \( L \) sites, governed by a nearest-neighbor hopping Hamiltonian with hopping strength $t_{\parallel}$ and coupled by a local hopping $t_{\perp}$. One of the chains is the system $S$, while the other is the ancilla $A$ used to extend the Hilbert space. This is the same model employed in \cite{tsitsishviliMeasurementInducedTransitions2024b,muzziEntanglementEnhancementInduced2025} to simulate non-Markovian dynamics.
\begin{equation}
\label{eq:Hmodel}
    H=-t_{\parallel}\sum_{i,\sigma}\hat c^\dagger_{i,\sigma}\hat c_{i+1,\sigma}-t_\perp\sum_i\hat c^\dagger_{i,1}\hat c_{i,2}+\text{h.c.}
\end{equation}

The ancilla is subject to a dephasing dynamics through Lindblad operators $L_i=\sqrt{\gamma}\hat n_{i,2}$, with $\hat n_{i,2}=\hat c^\dagger_{i,2}\hat c_{i,2}$ and $\gamma$ the dissipation rate. The system is initialized in two orthogonal Gaussian states $\rho_p$ and $\rho_q$ for the evaluation of \( d_2(\rho_p(t),\rho_q(t) \).

Since the dynamics is number conserving, we can actually express $d_2$ directly in terms of the correlation matrix $C$:
\begin{gather}
\label{eq:dqp_C}
\Tr(|\rho_p-\rho_q|^2)=\frac1{\Ntr^2}\sum_{\alpha,\beta}[\mathcal{D}(C_p^{\alpha},C_p^{\beta})+\\
\notag+\mathcal{D}(C_q^{\alpha},C_q^{\beta})-2\mathcal{D}(C_p^{\alpha},C_q^{\beta})];\\
\notag\mathcal{D}(C,C')\equiv\text{det}(\mathbb1-C-C'+2CC').
\end{gather}
The detailed derivation of this formula is reported in Appendix \ref{Appendix:d2C}.

We compare results obtained from both methods for chains of size \( L = 2, 3, 4, 5, 6\), with $t_\perp=t_\parallel=1$, $\gamma=1$, using \( N_{\mathrm{tr}} = 500 \) quantum trajectories in the Gaussian approach and a time discretization of \( \Delta t = 0.02 \) over a total evolution time of \( t_{evol} = 10 \). %As shown in Fig.~\ref{fig:blp_comparison}, both the Hilbert-Schmidt distance and the second-Rényi mutual information display non-monotonic behavior over time, indicating temporary revivals of distinguishability and correlations—clear signatures of non-Markovianity.

%For \( L = 4 \) and \( L = 6 \), we find excellent agreement between the results obtained via the Gaussian correlation-based protocol and the full density matrix evolution using QuTiP. 
As showed in Fig. \ref{fig:blp_comparison}, already for \( L = 7 \), the full simulation becomes unfeasible due to the exponential growth of the Hilbert space dimension (i.e., \( \sim 2^{2L} \) for the density matrix), and requires memory resources far beyond those available for standard computing hardware. Moreover, the computational time required by the QuTiP simulation scales exponentially with system size, whereas the Gaussian trajectory simulation remains tractable, completing in around one hour.

The computational complexity of our Gaussian protocol scales polynomially as \( \mathcal{O}(N_{\mathrm{tr}}^2 L^{2.38}) \), as obtained from a fit of larger system sizes in the range $L=16$ to $L=256$. Compared to the exponential scaling of exact methods, the Gaussian protocol exhibits a substantial speed-up, which enables the evaluation of standard non-Markovianity measures in large systems, well beyond the reach of brute-force methods.

\begin{table}[t]
    \centering
    \captionsetup{justification=justified}
    \begin{tabular}{c|c|c}
        Full state & QuTr+full & Gaussian \\
        %\vspace*{0.25cm}
        \hline
        %\vspace*{0.5cm}
        & & \\
        $\sim\mathcal{O}(2^{2L})$ & $\sim\mathcal{O}(\Ntr2^{L})$ & $\sim\mathcal{O}(\Ntr^2L^{2.38})$\\
        & &
    \end{tabular}
    \caption{Summary of the computational complexity $T_{\mathrm{comp}}$ required by different numerical methods. In the first column the evolution of the full state density matrix $\rho$. In the second column the complexity required by simulating the evolution of a full density matrix via the quantum trajectories (QuTr) approach; this method is not simulated in Fig. \ref{fig:blp_comparison}. In the third column, the complexity required by simulating Gaussian states.}
    \label{tab:my_label}
\end{table}

\section{Conclusions}\label{sec:Conclusions}

In this work, we have developed a new computational framework to evaluate non-Markovianity measures and estimate memory effects in free fermionic systems. Our method leverages the Gaussian nature of these systems to efficiently calculate two-points correlation functions and then non-Markovianity, reducing the computational cost from exponential to polynomial in system size compared to existing methods. This reduction has been tested in a simple toy-model, showcasing the great advantage of our method for large system sizes.

More in detail, we considered non-Markovianity in open fermionic quantum systems, whose dynamics preserves Gaussianity. Our computational protocol is based on the fact that a Gaussian state is fully described by two-point correlation functions, whose numerical simulation is very efficient. By recasting the BLP and LFS non-Markovianity measures in terms of the Hilbert-Schmidt distance and second-order Rényi mutual information, respectively, we obtained expressions that are computable from two-point correlation matrices. This allows for a substantial reduction in computational complexity compared to full density matrix simulations.

We compared the efficiency of our method against full density matrix simulations using QuTiP. The full state simulation quickly becomes intractable for $L \gtrsim 6$ due to exponential scaling, while the Gaussian approach remains efficient, with a computational time that scales polynomially with system size, thus exhibiting an exponential speed-up of the numerical protocol. This highlights the scalability of our method and its applicability to systems beyond the reach of exact methods.

Our approach exploits the Gaussianity-preserving nature of certain Lindbladian evolutions and the efficiency of quantum trajectory methods. We demonstrated that, even when the average state of the system evolves into a non-Gaussian state as a result of averaging over quantum trajectories, the relevant quantities can still be accurately computed from the correlation matrices of individual trajectories. Our method still presents some limitations, as it applies to non-interacting fermions with Gaussian-preserving Hamiltonians and Lindblad operators, and cannot be used for interacting models (e.g. Fermi-Hubbard systems), where two-points correlation functions are not sufficient anymore to describe the system. Therefore, our method can be employed mainly within theoretical models or engineered systems, an example being cold atoms in optical lattices.

Nonetheless, our work opens the door to practical evaluations of non-Markovianity in large-scale free-fermionic systems, including those undergoing measurements or coupled to structured environments. Future developments may extend this framework to other quantities such as the fidelity of noisy dynamics or to interacting systems -- e.g. random Clifford circuits.

\section*{Data Availability Statement} The data that support the findings of this study are available from the corresponding authors upon reasonable request.

\section*{Acknowledgements} The authors declare no competing financial interests. G.C. is supported by ICSC – Centro Nazionale di Ricerca in High-Performance Computing, Big Data and Quantum Computing under project E63C22001000006. G.C. acknowledges the CINECA award under the ISCRA initiative, for the availability of high performance computing resources and support.

\bibliographystyle{sn-aps}
\bibliography{GaussianMeasures}

\begin{appendices}

\onecolumn

\section{Proof of the contractivity of the Hilbert-Schmidt norm}\label{Appendix:d2}

We need to use the generalization of the Cauchy Schwartz inequality to operators $A$and $B$ subject to a completely positive map $\Phi$ \cite{ruskaiStrongSubadditivityImproved1994}:
\begin{equation}
\Phi(A^\dagger B)\Phi(B^\dagger A)\leq\Phi(A^\dagger A)\Phi(B^\dagger B)
\end{equation}

By choosing $A=\rho_p(t)-\rho_q(t)$ and $B$ equal to the identity, we find that
\begin{gather}
Tr[\Phi(\rho_p(t)-\rho_q(t))\Phi(\rho_p(t)-\rho_q(t))\leq\Phi(|\rho_p(t)-\rho_q(t)|^2);\\ d_2(\Phi(\rho_p(t)),\Phi(\rho_q(t)))\leq d_2(\rho_p(t),\rho_q(t))
\end{gather}

which proves that the distance $d_2$ contracts under CPTP maps and can thus be used for non-Markovianity measures.

\section{Derivation of the formula for $d_2$ in Gaussian systems}\label{Appendix:d2C}

In order to calculate $d_2$, we need to calculate the spectrum of $W$. Since $W$ is related to $\Gamma$ and $C$, we want to relate the spectrum of $W$ to the spectrum of $C$.

We start by observing that given a matrix $A$ with spectrum $\text{Spec}(A)=\{a_i\}$, then $\text{Spec}(A\otimes\begin{pmatrix}
    0&i\\-i&0
\end{pmatrix})=\{\pm a_i\}$. Therefore we find
\begin{gather}
\Spc(C)=\{c_k\}; \qquad \Spc(2C-1)=\{2c_k-1\}; \qquad \Spc(\Gamma)=\{\lambda_\Gamma\}=\{\pm(2c_k-1)\};\\
\Spc(1\pm\Gamma)=\{2c_k,2(1-c_k)\};\qquad \text{det}(1\pm\Gamma)=\prod_k4c_k(1-c_k)
\end{gather}

The partition function $Z(W)$ is given by Eq. \eqref{eq:ZexpW}:
\begin{gather}
Z(W)=\prod_{\lambda_w}\sqrt{2\cosh(\lambda_w/2)}=\prod_{\lambda_w}\sqrt{e^{\lambda_w/2}+e^{-\lambda_w/2}}
\end{gather}

We use that $e^W=\frac{1+\Gamma}{1-\Gamma}$, so that $e^{W/2}=\sqrt{\frac{1+\Gamma}{1-\Gamma}}$ and $\Spc(e^{W/2})=\Spc(\sqrt{\frac{1+\Gamma}{1-\Gamma}})=\{\sqrt\frac{1+\lambda_\Gamma}{1-\lambda_\Gamma}\}$. Therefore
\begin{gather}
Z(W)=\sqrt{\prod_{\lambda_\Gamma}\left(\sqrt\frac{1+\lambda_\Gamma}{1-\lambda_\Gamma}+\sqrt\frac{1-\lambda_\Gamma}{1+\lambda_\Gamma}\right)}=\sqrt{\prod_{\lambda_\Gamma}\frac2{\sqrt{1-\lambda_\Gamma^2}}}=\sqrt{\left(\prod_k\frac2{\sqrt{1-(2c_k-1)^2}}\right)^2}\\
Z(W)=\prod_k\frac1{\sqrt{c_k(1-c_k)}}=\left[\text{det}\left(\frac{1\pm\Gamma}2\right)\right]^{-1/2}
\end{gather}

We now want to calculate $Z(W'')=Z(W\oplus W')$, where $e^{W''}=\frac{1+\Gamma''}{1-\Gamma''}=e^We^{W'}=\frac{1+\Gamma}{1-\Gamma}\frac{1+\Gamma'}{1-\Gamma'}$

We can write
\begin{gather}
Z(W'')^4=\prod_{\lambda_{\Gamma''}}\frac{4}{1-\lambda_{\Gamma''}^2}=\text{det}\frac{4}{1-\Gamma''^2};\\
\frac{1\pm\Gamma''}{2}=\frac12\left(1\pm\frac{e^{W''}-1}{e^{W''}+1}\right);\qquad \frac{1-\Gamma''^2}{4}=\frac{1}{4}\left(1-\left(\frac{e^{W''}-1}{e^{W''}+1}\right)^2\right)=\frac{e^{W''}}{(e^{W''}+1)^2};\\
Z(W'')^4=\frac{\text{det}(e^{W''}+1)^2}{\text{det}(e^{W''})}; \qquad \text{det}(e^{W''})=\text{det}\left(\frac{1+\Gamma}{1-\Gamma}\frac{1+\Gamma'}{1-\Gamma'}\right)=\frac{\text{det}(1+\Gamma)}{\text{det}(1-\Gamma)}\frac{\text{det}(1+\Gamma')}{\text{det}(1-\Gamma')}=1;\\
e^{W''}+1=(1-\Gamma)^{-1}[(1+\Gamma)(1+\Gamma')](1-\Gamma')^{-1}+1=\\
=(1-\Gamma)^{-1}[(1+\Gamma)(1+\Gamma')+(1-\Gamma)(1-\Gamma')](1-\Gamma')^{-1}=2(1-\Gamma)^{-1}[1+\Gamma\Gamma'](1-\Gamma')^{-1}.
\end{gather}
\begin{gather}
Z(W'')^2=\frac{\text{det}[2(1+\Gamma\Gamma')]}{\text{det}(1+\Gamma)\text{det}(1+\Gamma')}=Z(W)^2Z(W')^2\text{det}\left(\frac{1+\Gamma\Gamma'}{2}\right);\\
\frac{Z(W\oplus W')}{Z(W)Z(W')}=\sqrt{\text{det}\left(\frac{1+\Gamma\Gamma'}{2}\right)};\\
\Gamma\Gamma'+1=[(2C-1)(2C'-1)+1]\otimes\mathbb{1}_2=2(1-C-C'+2CC')\otimes\mathbb{1}_2;\\
\text{det}\left(\frac{1+\Gamma\Gamma'}{2}\right)=\text{det}[(1-C-C'+2CC')\otimes\mathbb{1}_2]=\text{det}[(1-C-C'+2CC')]^2;\\
\frac{Z(W\oplus W')}{Z(W)Z(W')}=\text{det}(1-C-C'+2CC')
\end{gather}

This proves Eq. \eqref{eq:dqp_C}.

\end{appendices}

\twocolumn

\end{document}